\documentclass[lettersize,journal]{IEEEtran}
\usepackage{amsmath,amsfonts}
\usepackage{algorithmic}
\usepackage{algorithm}
\usepackage{array}
\usepackage[caption=false,font=normalsize,labelfont=sf,textfont=sf]{subfig}
\usepackage{textcomp}
\usepackage{stfloats}
\usepackage{url}
\usepackage{verbatim}
\usepackage{graphicx}
\usepackage{cite}
\usepackage{xcolor}
\usepackage{pgfplotstable}
\usepackage{diagbox}
\usepackage{makebox}
\usepackage{multirow}
\usepackage{makecell}
\usepackage{booktabs}
\begin{document}

\title{Anomaly Detection in Industrial Control Systems Based on Cross-Domain Representation Learning}

\author{Dongyang Zhan*,~\IEEEmembership{Member,~IEEE, }
        Wenqi Zhang,
        Lin Ye,
        Xiangzhan Yu, %~\IEEEmembership{Fellow,~OSA,}
        Hongli Zhang, %,~\IEEEmembership{Life~Fellow,~IEEE}% <-this % stops a space
        and~Zheng He

\IEEEcompsocitemizethanks{\IEEEcompsocthanksitem D. Zhan, W. Zhang, L. Ye, X. Yu, H. Zhang are with the School of Cyberspace Science, Harbin Institute of Technology, Harbin,
Heilongjiang, 150001.\\
Z. He is with the Heilongjiang Meteorological Bureau, Harbin, Heilongjiang, 150001.
\protect\\
E-mail: \{zhandy, 23S003108, hityelin, yuxiangzhan, zhanghongli\}@hit.edu.cn
%\IEEEcompsocthanksitem J. Doe and J. Doe are with Anonymous University.
\IEEEcompsocthanksitem * Corresponding Author: zhandy@hit.edu.cn}% <-this % stops an unwanted space
        
%\thanks{This paper was produced by the IEEE Publication Technology Group. They are in Piscataway, NJ.}% <-this % stops a space
%\thanks{Manuscript received April 19, 2021; revised August 16, 2021.}}


}
\maketitle

\begin{abstract}
Industrial control systems (ICSs) are widely used in industry, and their security and stability are very important. Once the ICS is attacked, it may cause serious damage. Therefore, it is very important to detect anomalies in ICSs. ICS can monitor and manage physical devices remotely using communication networks. The existing anomaly detection approaches mainly focus on analyzing the security of network traffic or sensor data. However, the behaviors of different domains (e.g., network traffic and sensor physical status) of ICSs are correlated, so it is difficult to comprehensively identify anomalies by analyzing only a single domain. In this paper, an anomaly detection approach based on cross-domain representation learning in ICSs is proposed, which can learn the joint features of multi-domain behaviors and detect anomalies within different domains. After constructing a cross-domain graph that can represent the behaviors of multiple domains in ICSs, our approach can learn the joint features of them by leveraging graph neural networks. Since anomalies behave differently in different domains, we leverage a multi-task learning approach to identify anomalies in different domains separately and perform joint training. The experimental results show that the performance of our approach is better than existing approaches for identifying anomalies in ICSs.
\end{abstract}


\section{Introduction}
\IEEEPARstart{I}{ndustrial} control systems (ICSs) are critical components of many modern industrial and infrastructure systems, such as power grids, and manufacturing plants. They enable remote monitoring and control of physical sensors or actuators based on network communication, which is very important for industry. 

However, ICSs are  vulnerable to cyberattacks, and the consequences of an attack can be serious. In recent years, lots of attacks against ICSs have triggered serious consequences. For instance, the Stuxnet worm \cite{karnouskos2011stuxnet} attacked Iranian nuclear facilities, and it demonstrated the potential for cyberattacks to cause physical damage to industrial systems. In 2021, the Colonial Pipeline was hit by a ransomware attack, causing widespread fuel shortages \cite{tsvetanov2021effect}. In 2015 and 2016, Ukraine's power grid was attacked, which caused power outages \cite{case2016analysis}. In the past, ICSs are isolated from the Internet and run in physically secure locations, so the security mechanisms are not well designed \cite{cook2017assessment}. Currently, as ICSs are increasingly connected to the Internet, ICSs are facing more and more security threats. Therefore, it is crucial to secure ICSs and ensure the security and reliability of critical infrastructure systems.

There are several challenges of securing ICSs. First, many legacy protocols are not designed with modern security features, making them vulnerable to attacks \cite{taylor2017security} such as denial of service, man in the middle, etc. For instance, MQTT is a widely used messaging protocol used for machine-to-machine communication in IoT networks that has been found to have many vulnerabilities \cite{ dinculeanua2019vulnerabilities}. Second, the devices (e.g., sensors and actuators) are mostly low-power and have limited computing power for security analysis and protection functions \cite{yang2023dependable}. Therefore, it is difficult for administrators to enhance security directly based on existing ICSs.

Many approaches have been proposed to secure ICSs. First, intrusion detection systems (IDSs) \cite{liao2013intrusion} can be applied to secure ICSs, which typically identify the network behavioral characteristics of known attacks but face the problems of identifying only few known attacks and the inability to detect unknown attacks. In addition, an important difference between ICS anomaly detection and network anomaly detection is that ICS managers have access to data from each sensor in the network (e.g., pressure, flow, etc.), based on which anomalies can be more accurately identified. This is because some anomalies do not cause changes in network traffic, but rather in anomalies in sensor data. 

Second, security analysis approaches based on behavior models are proposed, which construct the behavior/data models of every sensor, actuator and the interaction between them. If a physical sensor value or network packet does not fit the model, an anomaly alert is generated. But it is difficult to build comprehensive models of complex industrial processes systematically, so these approaches lack generality and are not widely used in ICSs \cite{kravchik2021efficient}.

\IEEEpubidadjcol

To address the limitations of traditional approaches, deep learning is utilized for anomaly detection in ICSs. Utilizing the capabilities of deep neural networks, these techniques can learn features that determine anomaly scores, resulting in improved accuracy in anomaly detection. Specifically, the Recurrent Neural Network (RNN) \cite{mandic2001recurrent} offers the potential to model the time series values of ICS physical devices, such as temperature sensors and pressure sensors. It captures the dependencies between inputs and outputs by extracting partial sequences from relevant inputs and scores anomalies based on the predicted or reconstructed features of ICS physical device values. Further building on the foundations of RNN, attention mechanism-imbued time series prediction models such as LSTM-NDT \cite{hundman2018detecting} and openGauss \cite{li2021opengauss} have been introduced, showcasing superior performance in identifying enduring correlations in ICS physical device values. Models such as TranAD \cite{tuli2022tranad} and Anomaly Transformer \cite{xu2021anomaly} adapt the Transformer architecture for time-series anomaly detection, which effectively capture rich associations across entire sequences using their self-attention weight distribution. Besides, the Generative Adversarial Network (GAN) \cite{creswell2018generative} also stands out for its prowess in multivariate anomaly detection through bias reconstruction, as demonstrated in models such as TAnoGAN \cite{bashar2020tanogan} and GAN-AD \cite{li2018anomaly}. However, it is difficult for these approaches to detect anomaly points in the time series values of ICS physical devices because they cannot model the topological and interaction relationships between nodes in ICSs.

Graph Neural Networks (GNNs) \cite{scarselli2008graph} have shown significant promise in graph representation learning. Graph Convolutional Network (GCN) \cite{kipf2016semi} is a convolutional neural network-based model that generates node embeddings by applying convolutional operations on graphs. Consequently, GCN performs well in graph embeddings and can incorporate inter-node correlations into the learning modeling process. Based on GCN, the Graph Deviation Network (GDN) \cite{deng2021graph} utilizes embedding vectors to capture the physical device features of ICSs (i.e., the values of each sensor or actuator in the real world). It encodes and learns the relationships between pairs of devices as graph edges. It predicts future device behaviors by applying an attention function to the neighboring sensors in the graph. Additionally, it identifies and explains deviations from the learned sensor relationships. While hybrid models (NSIBF \cite{feng2021time} and MTAD-GAT \cite{zhao2020multivariate}) use self-supervised methods to explicitly capture the correlation between features of different physical devices in time series data. Feng and Tian utilize Bayesian filtering to handle implicit relationships, while Zhao et al.'s approach involves direct learning through graph attention networks. However, both methods result in a high level of task complexity. GLIN \cite{shuaiyi2022global} analyzes the joint representations of local nodes and the global network to perform anomaly detection. This method is used for node-level anomaly detection within industrial control systems. MMGAN \cite{pandeva2019mmgan} jointly learns the features of physical device time series in both the time and frequency aspects, enhancing the model's ability to model distribution. However, these methods are designed for learning features of single-domain data (e.g., physical or network data) in the ICS without considering the joint modeling of multi-domain data. This leaves the models with inadequate detection capability.

In this paper, an anomaly detection approach based on cross-domain representation learning is proposed for ICSs, which integrates the states of both network and physical domains to learn and balance the multi-domain features of ICS elements (e.g., sensors) and detect anomalies in ICSs through predicted values on multiple domains. Since the operation status of ICSs can be reflected in different domains (i.e., the network and physical domains), our approach enables a more comprehensive analysis of ICS data by combining complementary information from different coarse-grained domains. To mitigate the sparsity problem of ICS behavior data in different domains, we propose a cross-domain graph construction method that presents the relationships between ICS elements in different domains in only one graph.

Our anomaly detection model comprises two stages. In the first stage, an attention-based GCN is used to learn shared representations of different domains based on comprehensive node features. In the second stage, our model learns domain-specific representations to capture the behavioral characteristics of ICSs in different domains. To accomplish this, shared representations are input into several graph models, each of which learns behavioral features in a different domain. Models in different fields predict the future behaviors of ICSs based on reconstructed graph structures. Anomaly detection is performed by calculating the difference between the predicted and actual states, and losses in different domains are used for learning the shared-domain and multi-domain representations. Additionally, we use a multi-gradient descent optimization algorithm based on multi-task learning to adaptively allocate the weight distribution between multiple tasks, which leads to better training results. Our model extracts high-dimensional features related to multi-domain behaviors in ICSs, which is difficult for other existing works.

In summary, the main contributions are presented as follows:

\begin{itemize}
\item A cross-domain graph representation method for ICSs is proposed to describe the behaviors between ICS elements of different domains in a single graph, which can represent the behavior features of ICS elements in different domains within a unique graph structure.
\item A novel anomaly detection model for ICSs based on attention-based graph convolutional networks is proposed, which learns both domain-shared and domain-specific behaviors to perform anomaly detection in multi-domains. To the best of our knowledge, this is the first work to jointly analyze multi-domain data for anomaly detection in ICSs.
\item After implementing the prototype, we conduct baselines and ablation experiments on the water treatment plant datasets with ground truth anomalies, which contains multi-dimensional data. Our results demonstrate that our model detects anomalies more accurately than baseline approaches.
\item The code for our proposed model is released on GitHub (\url{https://github.com/WenqiZhang-HIT/MGDN-project}) to enable interested researchers to reproduce and extend our work. 
\end{itemize}

The rest of this paper is organized as follows. Section \ref{s:related-work} summarizes the related work. The motivation and challenges of this paper are presented in Section \ref{s:motivation}. Section \ref{s:design} describes the proposed anomaly detection approach. The evaluation of our approach is performed in Section \ref{s:evaluation}. Section \ref{s:conclusion} summarizes this paper. 

\section{Related Work}\label{s:related-work}

In this section, we briefly summarize the relevant work on ICS anomaly detection. Furthermore, we briefly introduced previous work closely related to our research, focusing on methods of selecting different ICS features for learning.

\subsection{Statistics-based ICS Anomaly Detection}
Statistics-based methods have become a cornerstone for identifying irregularities that deviate from established behavioral patterns. One seminal approach is the application of multivariate statistical analysis. For example, a real-world application was demonstrated by Chiang, Russell, and Braatz \cite{chiang2000fault}, who employed the Multivariate State Estimation Technique (MSET) to model the normal operational regime of a chemical process, which is analogous to a nuclear plant's control system. The strength of their method lies in its ability to accommodate the nonlinear and complex interactions between various control loop parameters. Another significant contribution is the use of Statistical Process Control (SPC) \cite{caulcutt1996statistical}, particularly the CUSUM (Cumulative Sum) control chart, which has been skillfully applied to ICSs. For instance, Valle, Li, and Qin \cite{valle1999selection} demonstrated the utility of CUSUM in monitoring semiconductor manufacturing processes, which can be adapted to the context of ICSs for monitoring and signaling slight, yet significant, shifts in system performance that precede identifiable anomalies. Additionally, Time Series Analysis is a widely respected statistical method for detecting anomalies in ICS, thanks to its ability to capture temporal dependencies. A noteworthy implementation of this technique is provided by Basseville and Nikiforov \cite{basseville1993detection}, whose work on the detection of abrupt changes in signals and processes can be applied to SCADA systems \cite{daneels1999scada}. They discuss the use of Autoregressive Integrated Moving Average (ARIMA) models \cite{saboia1977autoregressive} among other methods, which are adept at identifying subtle anomalies in sensor data over time.

\subsection{Machine Learning-based ICS Anomaly Detection}
In contrast to traditional statistical methods, machine learning-based anomaly detection offers a data-driven approach for identifying irregularities within ICSs. These techniques are particularly skilled at handling high-dimensional data and complex pattern recognition tasks. One prominent method is the use of supervised learning algorithms, such as Support Vector Machines (SVMs) \cite{hearst1998support}, which have been applied to power systems for detecting cyber-physical attacks. Notably, Zahid et al. \cite{zahid2019electricity} utilized SVMs to classify the states of an electrical grid with high accuracy. Another important approach in machine learning is unsupervised learning, which includes techniques such as k-means clustering and Principal Component Analysis (PCA)\cite{abdi2010principal}. These methods have been instrumental in identifying outliers without the need for labeled data. An exemplary study by Abokifa and Haddad et al. \cite{abokifa2017detection} demonstrated the use of PCA in parsing through multivariate physical device data from water distribution systems, effectively isolating instances of operational anomalies. Semi-supervised techniques, which operate with a limited set of labeled data, have also been employed. One such approach is the use of One-Class Classification, where the model learns only from the 'normal' class to determine if new data points fit within that learned distribution. Zhang et al. \cite{zhang2022two} applied a one-class SVM to vibration data from rotating machinery in an ICS environment, effectively identifying potential mechanical failures. Furthermore, reinforcement learning has been explored for anomaly detection in ICSs. Algorithms have been designed to learn optimal actions through trial and error, enabling the system to dynamically adapt to new threats. Kim and Chayoung et al. \cite{kim2019designing} utilized a Q-learning algorithm to monitor network traffic in real-time, differentiating between regular operations and potential cyber security threats.

\subsection{Deep Learning-based ICS Anomaly Detection}

ICS anomaly detection methods based on deep learning utilize deep neural networks to predict multidimensional time series or reconstruct normal time series models. These methods then score anomalies based on the predicted or reconstructed bias. ICS is a complex system that encompasses data from various domains, including physical device display values measured in the real world, network transmission traffic, physical spatial topology, and more. Therefore, based on the dimensions of neural network learning, ICS anomaly detection methods based on deep learning can be categorized into two groups: single-dimensional and multi-dimensional.

The single-dimensional ICS anomaly detection method based on deep learning mainly refers to training and detecting data from only one dimension in ICSs. A framework for anomaly detection in univariate time series data is proposed by \cite{kao2019anomaly}. The framework utilizes statistical tests such as the Dickey-Fuller Test (FFT)\cite{mushtaq2011augmented}, and Pearson product moment correlation coefficients. It also incorporates different schemes of GRU's deep learning model to perform anomaly detection on various categories of univariate time series. In order to prevent new errors in the generative adversarial networks (GANs) \cite{goodfellow2020generative} for finding the optimal mapping from real-time space to potential space, the LSTM-based VAE-GAN \cite{niu2020lstm} trains encoders, generators, and discriminators together. This allows for the simultaneous utilization of the mapping abilities of encoders and discriminators, as well as the detection of anomalies in univariate time series based on differences in reconstruction and discriminant results. Aggarwal uses reconstruction errors based on Auto-Encoder (AE) \cite{hinton2011transforming} to measure single-dimensional anomalies. DAGMM \cite{zong2018deep} consists of a compression network and an estimation network, which combine the Deep Auto-Encoder (DAE) \cite{lange2010deep} density estimation processes for end-to-end joint training. In addition, many methods have applied graph neural networks to anomaly detection, such as GDN \cite {deng2021graph}, which combines structural learning with Graph Attention Network (GAT)\cite{velivckovic2017graph}. GDN applies attention mechanism to adjacent physical devices on the graph to learn the features of each timestamp of device value, and uses prediction error to detect anomalies in ICSs.

Unlike the single-dimensional methods described above, multidimensional ICS anomaly detection methods typically perform anomaly detection based on information from multiple dimensions of ICSs. For example, GGM-VAE \cite{guo2018multidimensional} considers the inherent multimodal distribution in time series data, uses a Gated Recurrent Unit (GRU) \cite{dey2017gate} to discover the correlation between time series, and then uses Gaussian mixture priors in latent space to characterize multimodal data. GLIN \cite{shuaiyi2022global} takes each physical device in ICSs as a node, introduces a fixed topology as a graph structure, and implements node-level anomaly detection by merging the local expressions of nodes and the global expressions of the network. Moshe Kravchik and Asaf Shabtai \cite{kravchik2018detecting} proposed an attack detection method based on simple lightweight neural networks, namely one-dimensional convolution and autoencoder, applied to the time and frequency aspects of time series data.

Although the above methods are referred to as deep learning-based multidimensional ICS anomaly detection, the analysis of data acquired by physical devices (e.g., sensors) is not considered multi-domain analysis. Even though the time series in ICSs are analyzed in terms of time and frequency aspects, the data are obtained from one domain. Some methods introduce information from other dimensions (e.g., the fixed spatial topology of the devices in ICSs) as fixed known information to assist in the learning of the main dimension information, but the introduced information does not represent the behavior features of another domain. Therefore, these methods are often referred to as "multi-variate", rather than "multi-domain". The method proposed in this paper elevates multidimensional ICS anomaly detection to a broader definition, namely multidomain-multivariate ICS anomaly detection. At the same time, information from multiple domains (physical device domain and network transmission domain) in ICSs is both considered. The information in each domain is also multi-variate. Each physical device in ICSs is considered a node, which includes the device value time series on the physical domain and the traffic feature time series on the network domain. The information from multiple domains is fused for cross-domain learning to deeply explore the inherent correlation and potential characteristics of the nodes in ICSs. Besides, multidimensional learning and anomaly detection are performed within each domain to further analyze the security of each specific domain.

\section{Motivation and Challenges}\label{s:motivation}
This section discusses the motivation for considering the cross-domain learning of multi-dimensional data in ICS anomaly detection as well as the challenges brought by multi-dimensional cross-domain representation learning.

\subsection{Motivation}

ICS is a complex system that comprises traditional industrial control systems as well as computer networks and communication technologies. Its behavior is typically manifested across multiple domains, such as physical device(e.g. sensor or actuator) values, network traffic, and more. Different behaviors exhibit varying patterns across these domains. For instance, ICSs can be targeted with attacks across different domains, such as disrupting devices in the physical domain or launching denial-of-service attacks on nodes in the network domain.

Simultaneously, for a given behavior, the responses across different domains can also differ significantly. For example, a particular attack may cause a significant fluctuation in sensor values in the physical domain while the network domain data remains normal. Alternatively, it may result in distinct anomalies across two different domains simultaneously \cite{feng2017multi}. Algorithms designed to focus on domain-specific features may be more sensitive to anomalies unique to those domains. Conversely, a broader selection of features may detect a wider range of abnormal patterns. For instance, an attacker can inject forged water level sensor data through network intrusion, causing the system to falsely perceive the water level as normal \cite{mr2021machine}. This leads to operational deviations without significant changes in the physical device data. Consequently, anomalies would be challenging to detect promptly if monitoring is confined solely to the physical domain. However, these anomalies can be detected by identifying unusual traffic patterns or undesired communication frequencies in the network domain.

Therefore, analyzing the behavioral characteristics of ICSs solely in a single domain (either physical or network) cannot perceive the security of the entire ICS. Utilizing single-domain feature learning for anomaly detection may result in a high rate of false negatives or false positives. For instance, several single-domain anomaly detection methods, such as DTAAD \cite{yu2023dtaad}, MSCRED \cite{zhang2019deep}, and OmniAnomaly \cite{su2019robust}, exhibit significant differences in precision and recall in the experiment, which means it is difficult for them to obtain balanced results for real-world anomaly detection. This issue is largely attributed to the data sparsity problem caused by single-domain learning \cite{manzhos2022advanced}.

The data sparsity problem was first introduced in recommendation systems. In recommendation systems, data sparsity refers to the scarcity of interaction data between users and items, making it challenging to accurately predict user ratings for items \cite{resnick1997recommender}. The problem of data sparsity not only occurs in recommendation systems but also manifests in various other domains. For instance, Ramchandran and Sangaiah \cite{ramchandran2018unsupervised} noted that in the field of anomaly detection, the scarcity of anomaly data results in highly sparse datasets. There are several approaches to address this issue, including improved feature engineering, enhanced algorithms, and the utilization of better datasets.

Building upon the potential data sparsity issues that may arise from single-domain ICS anomaly detection, a new approach for anomaly detection in ICS is proposed, which includes multi-domain feature extraction, data learning, and feature prediction.

\subsection{Challenges \& Solutions}

Joint modeling of multi-domain data faces significant challenges due to data heterogeneity, involving variations in data types, structures, and formats across domains. Physical data typically appears as continuous time series, whereas network data is often discrete and event-driven, complicating data fusion. Additionally, different domains may interpret the same event differently, necessitating effective data correlation. Furthermore, the complexity of multidimensional data challenges model design, with high-dimensional datasets potentially leading to sparse feature spaces that impair learning effectiveness \cite{manzhos2022advanced}. This requires sophisticated models to capture relationships across fields, necessitating advanced architectural and parameter adjustments. Finally, in multi-domain modeling, the cross-domain learning process in high-dimensional space can cause the curse of dimensionality phenomena, such as gradient explosion or gradient vanishing \cite{peng2023interpreting}. In this section, the three primary challenging and significant issues are detailed, and our proposed solutions are presented.

\subsubsection{How to extract multi-domain features from ICSs and transform each domain into a unified and efficient feature representation?}

In the real world, ICSs exhibit different behaviors across multiple domains, and the measurement and representation of data also vary. Therefore, a primary challenge is how to obtain processable data from ICSs across multiple domains and transform the data into a unified format suitable for learning.

To address this challenge, we first conduct a survey and identify the widely used SWaT dataset, which contains data from both the physical and network domains simultaneously. We examine the initial recording formats of the behavioral data in each domain and find the differences between them. Based on the observations, it is challenging to combine and unify the initial data from the physical and network domains. For instance, in the physical domain, which is often used for single-domain learning, the data directly records the physical values of each sensor in the ICSs at a per-second granularity. On the other hand, the network domain records the extracted results of all network packets transmitted within the ICS network over time. Thus, our first step is to align the data at a consistent time granularity and extract the network domain data into feature values for each node, facilitating the acquisition of a multi-domain feature matrix for the ICS.

\subsubsection{How to efficiently and comprehensively represent node features across multiple domains in ICSs?}

To perform multi-domain analysis, it is necessary to efficiently represent the node time series features of different domains, and behavioral sequence data and topological relationship data should be preserved, thus facilitating the application of various deep learning models for analysis.

To achieve this, we propose a novel representation method called multi-graph structure, which allows the representation of multi-domain data in ICSs on a single graph structure while preserving inter-domain node correlations and temporal information. Each single-domain feature is represented on an individual graph, where the edges represent the associations between two nodes. In this graph, each node represents a sensor or an actuator, which has its own temporal feature vector. By mapping multiple single-domain graph structures based on nodes and concatenating the feature vectors of each node, a multi-graph representation can be obtained.

\subsubsection{How to learn the behavioral characteristics of ICSs from multiple domains associatively to achieve efficient anomaly detection?}

Unlike single-domain learning models, to capture additional cross-domain correlation information, it is necessary to design and implement an efficient joint feature learning model that can adequately analyze the multi-graph structure and utilize the learned results for practical feature prediction across different domains.

To that end, we first conduct joint learning on the multi-graph structure to learn the cross-domain features of ICSs on multiple domains. Furthermore, to achieve more accurate anomaly detection, we learn the node features of each specific domain based on the results of cross-domain learning. After that, multi-domain anomaly detection is performed by predicting the node behaviors on each domain.

As a result, the model follows a multi-task learning mechanism. However, due to the competitive and interdependent nature of the parent-feeding relationships among the multiple convolutional layers in the child, it is challenging to address the resulting multi-objective optimization problem using traditional methods. Therefore, during the forward optimization process, a multi-gradient descent optimizer is introduced to dynamically balance the weights of the multiple domain-specific convolutional layers, accelerating model convergence and avoiding issues such as gradient explosion.

\section{Methodology}\label{s:design}

\begin{figure}[!t]
    \centering
    \includegraphics[width=3.5in]{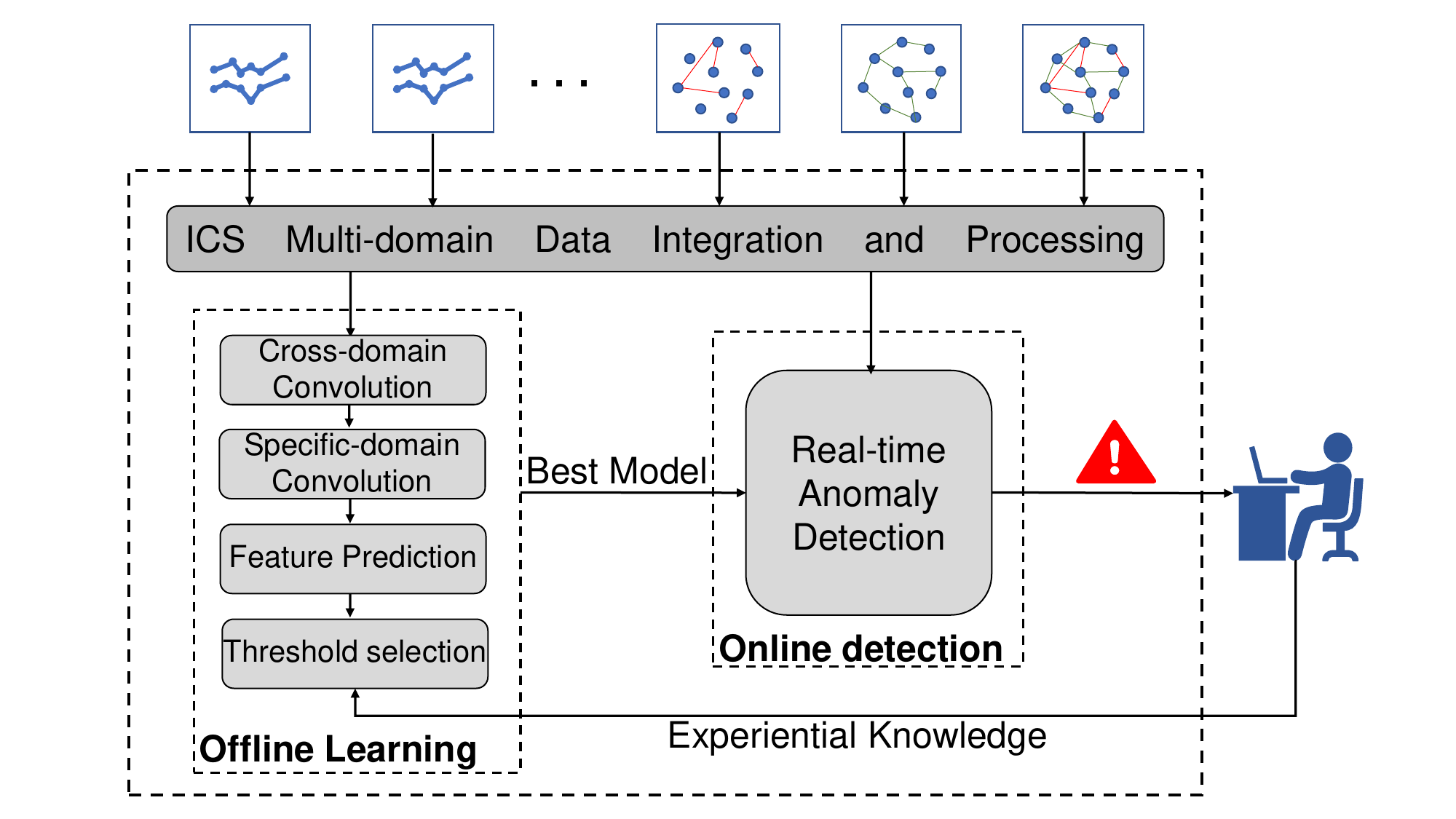}
    \caption{The construction process of multi-graph representation structure.}
    \label{fig:fig_1}
\end{figure}

To perform accurate anomaly detection in ICSs, a multi-dimensional cross-domain anomaly detection approach is proposed by integrating the above-mentioned solutions, consisting of the processing and representation of the multi-dimensional data in ICSs and cross-domain feature learning and prediction. The overall process of the proposed method is shown in Figure \ref{fig:fig_1}.

\subsection{Problem Definition}

In this paper, our training data consists of sensor data (i.e., multivariate time series) and network data from $N$ sensors over $T_{\mathrm{train}}$ time steps. The sensor data is denoted $\mathbf{s}_{\mathrm{train}}=\left[\mathbf{s}_{\mathrm{train}}^{(1)},\cdots,\mathbf{s}_{\mathrm{train}}^{(T_{\mathrm{train}})}\right]$. In each time step $t$, the sensor values $\mathbf{s}_{\mathrm{train}}^{(t)}\in\mathbb{R}^N$ form an $N$ dimensional vector representing the values of our $N$ sensors. Following the many unsupervised anomaly detection approaches, the training data consists of only normal data.

For each domain $d\left(d=\{1,...,D\}\right)$, we first construct an undirected weighted graph ${\cal G}_{d}=({\cal V},{\cal E}_{d})$. As these $D$ domains are correlated and share the same set of nodes, we then construct the cross-domain graph as an undirected weighted multi-graph ${\mathcal{G}}=({\mathcal{V}},{\mathcal{E}})$, which contains the node set $V$ with $N$ nodes and the edge set $E$ with $D$ types of edges, i.e., ${\mathcal{E}}=\{{\mathcal{E}}_{1},...,{\mathcal{E}}_{D}\}$.

Our problem can be formally stated as follows: with an undirected weighted multi-graph ${\mathcal{G}}=({\mathcal{V}},{\mathcal{E}})$, and the node feature matrix $\mathbf{X}\stackrel{.}{\in}\mathbb{R}^{N\times M}$ representing input for each node as an $M$ dimensional feature vector, our goal is to learn a set of embeddings for all nodes in each subgraph ${\cal G}_{d}$, i.e., $\mathcal{X}=\{\mathbf{X}_{1},...,\mathbf{X}_{D}\}$ ($\mathbf{X_{d}}\in\mathbb{R}^{N\times E}$ is the node embedding in subgraph ${\cal G}_{d}$), and predict the characteristics of the next time step $ \tau $ for each node. Test data for anomaly detection, which comes from the same $N$ sensors but over a set of $T_{\mathrm{test}}$ time steps, the test data is denoted $\mathbf{s}_{\mathrm{test}}=\left[\mathbf{s}_{\mathrm{test}}^{(1)},\cdots,\mathbf{s}_{\mathrm{test}}^{(T_{\mathrm{test}})}\right]$.

\subsection{Multi-Graph Construction}

As previously noted, the majority of current anomaly detection approaches for ICSs are constrained to analyze information from a single domain, such as physical domain time series (i.e., physical values of individual sensors over time), which makes it challenging to analyze the inherent correlations among nodes (i.e., sensors) in different domains. This limitation results in the absence of some key information, as the behaviors of ICSs are not limited to one domain. For instance, most ICSs are designed to obtain real-time physical data from sensors and send instructions to actuators through network traffic, so the physical data and network data are both important for security analysis.

A multi-graph structure is proposed to integrate and represent information from multiple dimensions of ICSs on a single graph, which can be used to learn the potential correlations and features of nodes across domains. To this end, we extract node embeddings from both the physical and network domains. 

In real-world scenarios, information from different domains is represented and measured in various ways. For instance, in the physical domain, sensor values can be directly recorded at each time step (in seconds), which can be easily converted into time series information. On the other hand, in the network domain, all transmitted network packets over a period of time are recorded, resulting in a large amount of data per unit time. This data cannot be directly used as time series information for nodes, and the analysis results of network packets cannot be directly used as network domain characteristics of each node. Although both domains measure information within the same time period, they cannot be directly mapped and fused because the time granularity differs. Therefore, simply merging the data from these two domains is difficult and meaningless, requiring mathematical and statistical methods to unify the data and map them to the same time granularity.

To overcome these challenges, we introduce an embedding vector for each sensor in each domain, representing its physical and network features: 
\begin{equation}
    \mathbf{X}_{\mathrm{phy}}^{(i)},\mathbf{X}_{\mathrm{net}}^{(i)}\in\mathbb{R}^T,\text{for}i\in\{1,2,\cdots,N\}
\end{equation}
This enables us to effectively present the underlying relationships among different domains of data.

We follow the methods of GDN \cite{deng2021graph} to extract and construct physical embeddings $\mathbf{X}_{\mathrm{phy}}^{(i)}$ based on physical sensor data. When dealing with network data, we analyze the unique characteristics of network traffic. Specifically, we calculate the number of data packets received, sent, and total data load of each node within a consistent time granularity (in seconds) that matches the physical information. To capture the temporal nature of this data, we extract the time series of these features through a sliding window. Then, the original network traffic data is simplified into a representative feature time series. Finally, the time granularity of network domain information is aligned with physical domain information to facilitate the merging of multi-domain information into a shared graph using time mapping methods.

\begin{figure}[!t]
    \centering
    \includegraphics[width=3.5in]{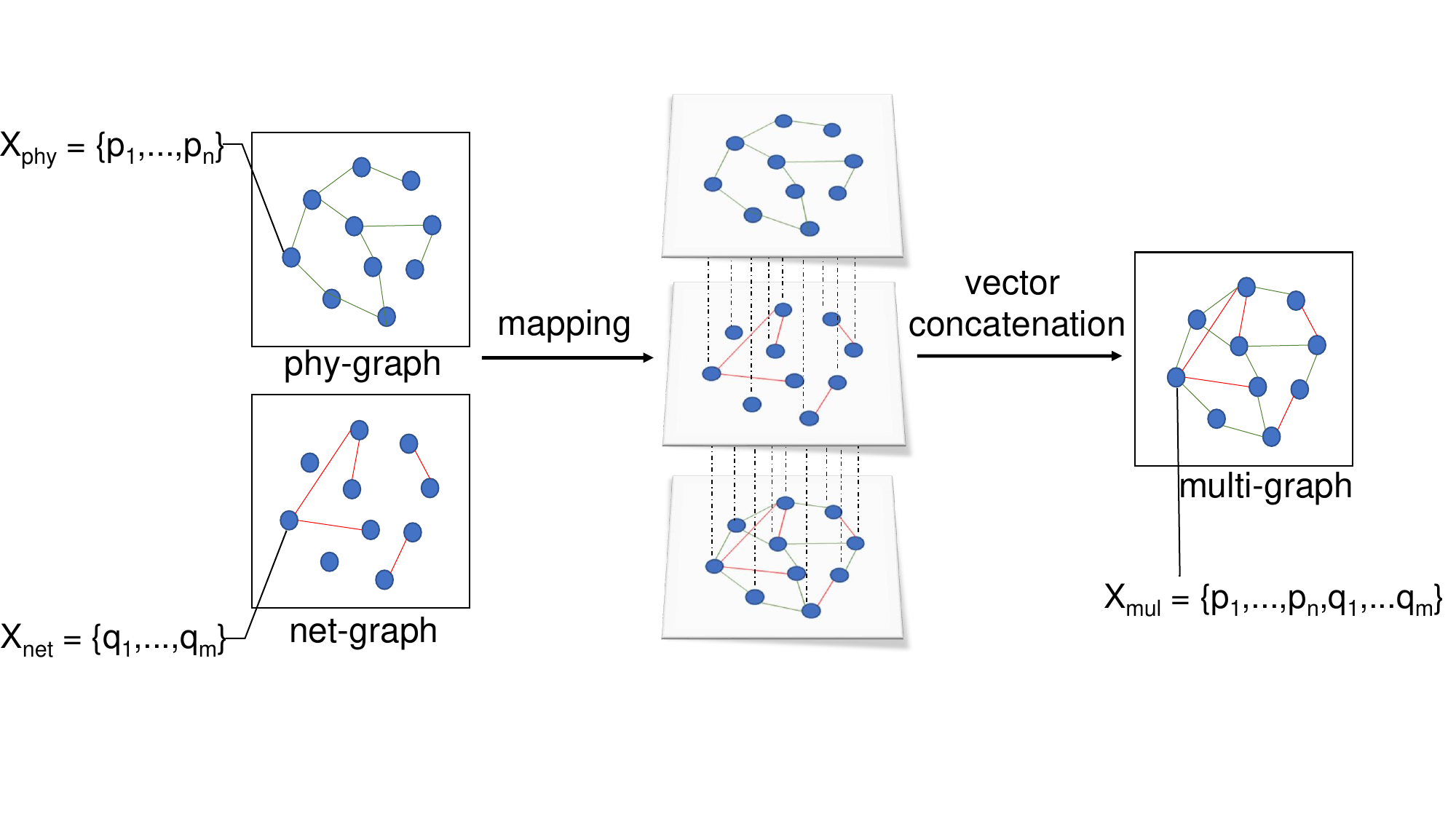}
    \caption{\centering{The construction process of multi-graph representation structure.}}
    \label{fig:fig_2}
\end{figure}

The multi-graph structure is constructed based on the results of the above approaches. The cosine calculation method is employed to calculate the probability of node correlation based on the node embedding matrix for each domain. Then, we apply the $Topk$ threshold to identify the highest-probability neighbors for each node, thus generating a graph structure for each domain. Finally, we consolidate the edges of different domains into a single graph, allowing for the possibility of multiple edges between any two nodes. This process can generate a multi-graph that effectively represents the complex relationships within the data. To form the node shared embedding in the multi-graph structure, we append the network domain embedding vector to the rear of the physical domain embedding vector through the vector concatenation method \cite{shuaiyi2022global}. The construction process of the multi-graph structure is illustrated in Figure \ref{fig:fig_2}.

\subsection{Attention-based Cross-Domain Graph Neural Network}%%

\begin{figure*}
\centering
\includegraphics[width=1\textwidth]{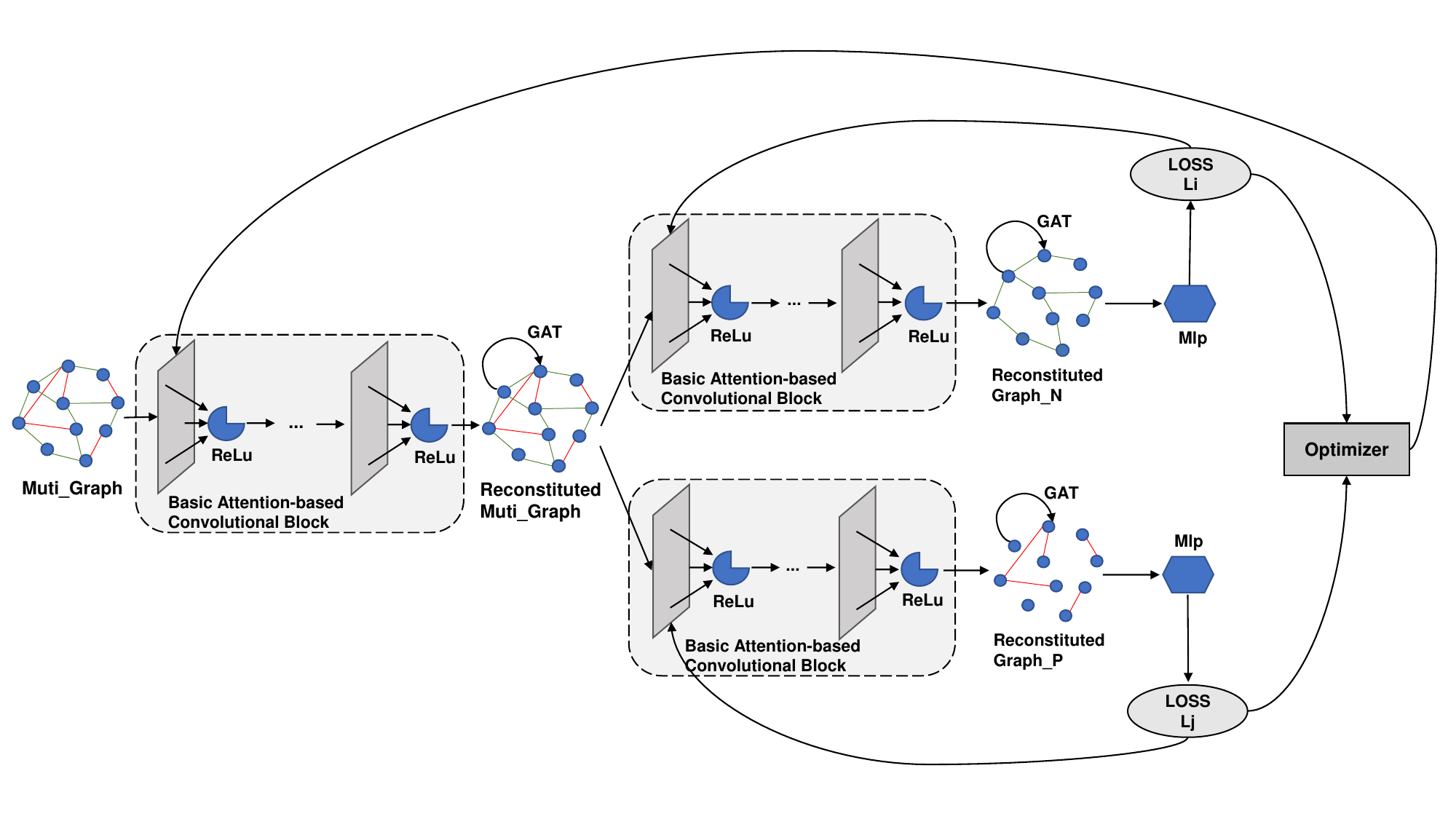}
\caption{\centering{The overall framework for cross-domain representation learning in our model.}}
\label{fig:fig_3}
\end{figure*}

Our model takes a multi-graph structure as input and first delves into modeling and exploring the correlation between dimensions in ICSs through domain-crossed convolutional layers. Then divide the learned results into separate domains for the second round of domain-specific learning. Finally, the multilayer perceptron is used in the prediction module to perform feature prediction based on the output of convolutional layers in each domain. Specifically, the convolution operations in the model are implemented by multiple basic attention-based convolutional blocks with different targets. The framework of the model is shown in Figure \ref{fig:fig_3}.

\noindent\textbf{Basic Attention-based Convolutional Block} \ To effectively learn node embeddings, we perform convolutional learning using basic blocks on the graph structure. The basic convolutional block consists of graph convolutional layers and incorporates attention mechanisms to better explore the potential correlations between nodes. The introduction of graph attention mechanisms allows for a more holistic consideration of the dynamic relationships between nodes on a global scale. 

To calculate the similarity between the embedding vectors of node $i$ and its candidates $j\in\mathcal{C}_i$ (in the case without prior information, the candidates of sensor $i$ are all sensors $\mathcal{C}_i$), the calculation formula is as follows:

\begin{equation}
    e_{ji}=\dfrac{\mathbf{x_i}^{\top}\mathbf{x_j}}{\|\mathbf{x_i}\|\cdot\|\mathbf{x_j}\|} \quad\mathrm{for}j\in\mathcal{C}_i
\end{equation}

Among them, $e_{ji}$ represents the normalized dot product between the embedding vectors of sensor $i$, and the candidate $j$. The first $k$ normalized dot products are selected by the fuction $FstK$ to form the adjacency matrix $A$. The value of $k$ can be selected based on expectations and prior knowledge.

\begin{equation}
 \mathcal{N}_i~~=~~\mathsf{F s t K}(\{e_{j i}:j\in{\mathcal{C}}_{i}\} 
\end{equation}

\begin{equation}
    \begin{cases}\mathbf{A}_{ji}=1 \ , \ j\in\mathcal{N}_i\ 
 \\ \mathbf{A}_{ji}=0 \ , \ j\notin\mathcal{N}_i\ \end{cases}
 \quad\mathrm{for} \ i\in\mathcal{V}
\end{equation}

Based on the adjacency matrix, the basic convolutional block in the model utilizes the node embedding ${X}_i$ and their neighboring embedding ${X}_j$ to compute the aggregated representation of each node. The calculation formula for feature extraction in the basic convolutional block is defined as follows:
\begin{equation}
X_{i}^{(l+1)}=\mathsf{ReLU}\left(\mathbf{W}\lambda_{i,i} X_{i}^{(l)}+\mathbf{W}\sum\limits_{j\in\mathcal{N}_{i}}\lambda_{i,j} X_{i}^{(l)}\right)
\end{equation}
$\mathbf{W}$ is a weight matrix that can be trained, $X_{i}^{l}$ is the embedding of node $i$ in layer $l$, and $X^{0}$ is the input feature matrix, $\lambda_{i,j}$ is the attention coefficient of node $j$ to node $i$, which is obtained by normalizing the attention score and the calculation formula is as follows, $\mathbf{a}$ is the learning coefficient vector of the attention mechanism:

\begin{equation}
\lambda_{i,j}=\mathsf{Softmax}\left(\mathsf{LeakyReLU}\left(\mathbf{a}^{\top}\left(X_{i}\oplus X_{j}\right)\right)\right)
\end{equation}

\noindent\textbf{Cross-Domain Learning} \  To learn multiple types of node representations on multiple domains, a cross-domain graph neural network is proposed, which mainly has two stages: multi-graph learning and subgraph learning. In the multi-graph learning phase, all domain information is reflected on the same graph, where each node behaves differently in different domains. By leveraging the interrelatedness of these representations, we aim to learn a shared representation that represents cross-domain shared information in multiple domains. In the subgraph learning stage, we focus on learning the specific representation of each node on each domain to encode specific information on different subgraphs. The detailed description of our methodology and the overall architecture of our model are illustrated in Figure \ref{fig:fig_3}.

Specifically, our model comprises basic convolutional blocks with multiple objectives, following the multi-task learning regime (MTL) \cite{ruder2017overview}. The cross-domain learning layer encodes multi-graph structures and node shared attributes as the input of the basic attention-based convolutional block to generate shared node embedding. Next, the learned shared node embeddings are split into node embeddings on two domains. The convolutional layers of two specific domains take the node embeddings on each subgraph as inputs and encode them to generate node embeddings on specific domains. The interaction structure of all graph convolutional layers is depicted in Figure \ref{fig:fig_3}.

As described above, our model consists of multiple graph convolutional layers, each with different learning contents. The parameter ${W}_{s}$ of cross-domain graph convolutional layer is shared by multiple domains, while the parameters ${W}_{d} ~~(d=1,...,D)$ of domain-specific graph convolutional layers are used for each specific domain. To train the model and benefit the learning process of all domains, that is, to enable multiple tasks in the model to achieve an optimal state, we need to optimize all target parameters $({W}_{S} ,~{W}_{1},~...,~{W}_{D})$.

In MTL \cite{ruder2017overview}, the commonly used method for optimizing the objective function is to calculate the weighted sum of the loss $Ld$ for all tasks statically or dynamically. However, stacking multiple layers brings additional difficulties to the model training, and adjusting the weights of each task to obtain the optimal training effect is time-consuming \cite{ouyang2019learning}. In our model, the optimization goal is to minimize the loss of each specific domain learning module, as shown in formula (7):

\begin{equation}
\begin{cases}\min\limits_{{W}_s,{W}_1}L_1({W}_s,{W}_1)\\...\\ \min\limits_{{W}_s,{W}_D}L_D({W}_s,{W}_D)\end{cases}
\end{equation}

To find the optimal solution for each objective, that is, the optimal solution in each domain, we employ a multi-gradient descent optimizer \cite{sener2018multi}. As in single-objective optimization, multi-objective optimization problems can be solved to local optima through gradient descent. The Multiple Gradient Descent Algorithm (MGDA) \cite{desideri2012multiple} utilizes the Karush–Kuhn–Tucker (KKT) conditions \cite{mangasarian1994nonlinear} (Eq. (8)), which is a necessary condition for the optimal solution of multi-objective optimization problems and is defined as follows:

\begin{equation}
\begin{cases}\sum\limits_{d=1}^D\alpha_d\dfrac{\partial L_d({W}_s,{W}_d)}{\partial{W}_s}=0\\ \dfrac{\partial L_d({W}_s{W}_d)}{\partial{W}_d}=0,(\forall d\in\{1,...,D\})\\ \sum\limits_{d=1}^D\alpha_d=1\\ \alpha_d\geq0,(\forall d\in\{1,...,D\})\end{cases}
\end{equation}
where $\alpha_d$ is the weight of objective $L_d({W}_s,{W}_d)$.

Either the solution to equation Eq. (10) is 0, ensuring that the result satisfies Eq. (8), or there exists a solution that provides a descent direction to enhance all tasks in Eq. (7) as mentioned in \cite{gordon2012karush}. Consequently, solving Eq. (8) is equivalent to solving Eq. (10), as illustrated below. Additionally, Eq. (9) signifies the weighted sum of gradients for each specific domain.%不通顺-----已修改

\begin{equation}
{K}=\left\|\sum_{d=1}^{D}\alpha_{d}\frac{\partial L_{d}(\Theta_{s},\Theta_{d})}{\partial\Theta_{s}}\right\|_{2}^{2}
\end{equation}

\begin{equation}
 \min\limits_{\alpha_{1},...,\alpha_{D}}\left\{~{K}~\Biggl|~\sum_{d=1}^D\alpha^d=1,~\alpha^d\geq0\right\}
\end{equation}

The resulting MTL algorithm enables gradient descent of multiple tasks on specific parameters, followed by solving Eq. (8) and applying the solution Eq. (9) as a gradient update to shared parameters. When the number of specific domains is 2, the optimization objective Eq. (9) can be simplified as:

\begin{equation}
\begin{split}
\min\limits_{\alpha}\left\|\alpha\dfrac{\partial L_1({W}_s,{W}_1)}{\partial{W}_s}+(1-\alpha)\dfrac{\partial L_2({W}_s,{W}_2)}{\partial{W}_s}\right\|_2^2\\s.t.~0\leq\alpha\leq1
\end{split}
\end{equation}

The Eq. (10) is a unary quadratic equation of $\alpha$. With the weight $\alpha$, we update the parameters of the model as follows:

\begin{equation}
    {W}_{d}={W}_{d}-\eta\dfrac{\partial L_{d}({W}_{s},{W}_{d})}{\partial{W}_{d}}
\end{equation}

\begin{equation}
    {W}_s={W}_s-\eta\sum\limits_{d=1}^D\alpha_d\dfrac{\partial L_d({W}_s,{W}_d)}{\partial{W}_s}
\end{equation}

\noindent\textbf{Feature Prediction} \ After the embedding learning, we obtain representations for all nodes on each specific domain, namely $\mathbf{X}_{d} = \{\mathbf{x}_{d,1}^{(t)},\cdots,\mathbf{x}_{d,N}^{(t)}\}$. For each node i, we compute the inner product of the feature representation $\mathbf{x}_{d,i}^{(t)}$ and the corresponding time series embedding $\mathbf{v}_i$, and employ the product of all nodes as the input of the Multilayer Perceptron (MLP) \cite{taud2018multilayer} with output dimension N to predict the node vector at time t (i.e. the value of the sensor):

\begin{equation}
\begin{split}
    y=&f_3(w_3f_2(w_2f_1(w_1[\mathbf{v}_1\circ\mathbf{x}_1,...,\mathbf{v}_N\circ\mathbf{x}_N]\\&+b_1)+b_2)+b_3)
\end{split}
\end{equation}

$\mathbf{v}_i$ is an embedded vector randomly initialized for each node and trained together with the model. The similarity between $\mathbf{v}_i$ and $\mathbf{v}_j$ represents the similarity of behavior between nodes $i$ and $j$. These embeddings allow attention mechanism to consider more association possibilities, which helps in learning graph structures.

A three-layer Perceptron is used as the output layer, where $f_1$, $f_2$
and $f_3$ are Linear, BatchNorm1d and LeakRelu, $w_{i}$ and $b_{i}$ are the weight matrix and offset vector of layer $i$, respectively, and $y$ is the final prediction vector in each dimension.

To efficient train our model, after obtaining the predicted output $\mathbf{\hat{p}^{(t)}}$ of the model at time $t$, we employ the MAE (Mean Absolute Error) between the predicted output and the actual observation data $\mathbf{{r}^{(t)}}$ as the loss function on each domain-specific convolutional layer:

\begin{equation}
L1loss=\frac{1}{T_{\mathrm{train}}-w}\sum_{t=w+1}^{T_{\mathrm{rain}}}\left\|\mathbf{{y}}^{(\mathbf{t})}-\mathbf{r}^{(\mathbf{t})}\right\|
\end{equation}

\section{Evaluation}\label{s:evaluation}
In this section, we first describe the experimental dataset and performance indicators. Then, we evaluate our model and compare it with several state-of-the-art baselines that are designed for detecting anomalies in ICSs. Finally, we analyze the experimental results to demonstrate the effectiveness of our model.

We use the PyTorch\cite{paszke2017automatic} 1.12.1 with CUDA 11.3 and PyTorch Geometric\cite{fey2019fast} 2.1.0 library to implement our model and train all models. We set the hyper parameters as described in the baseline models of the papers. We use the following hyperparameter values in our model.

\begin{itemize}
    \item Batch size = 32
    \item Window size = 15
    \item The value of K in the $FstK$ = 20
    \item Dropout in convolutional blocks = 0.2
    \item Depth of hidden layers in convolutional blocks = 2
    \item The momentum of the SGD optimizer = 0.9
\end{itemize}

The effect of different learning rate combinations for each task on anomaly detection performance is analyzed in Section \ref{subsec:results_discussion}. To accelerate the convergence process, we employ the Stochastic Gradient Descent (SGD) optimizer \cite{bottou2012stochastic} to guide model updates. To achieve faster and smoother optimization, we incorporate a momentum term into the SGD optimizer. Apart from the learning rates, other hyperparameters are selected using grid search and empirical rules.
%参数设置的讨论、展示
%回答问题一（3）和二（5）
\subsection{Dataset}

By adjusting the data preprocessing algorithm, our approach can be readily extended to various dual-domain ICS datasets. By augmenting the number of single graphs in the multi-graph representation algorithm, expanding domain-specific learning modules, and optimizing hyperparameters for the multi-gradient descent optimization algorithm, our method can be applicable to anomaly detection across three or more domains within ICS. It is also adaptable to various types of ICS or other industrial environments. However, after extensive research and investigation, show that many existing public ICS anomaly detection datasets only offer physical domain-related data (i.e., sensor values) \cite{ahmed2017wadi,al2020intrusion,downs1993plant}, or only provide traffic data in the network domain \cite{choi2019comparison,dehlaghi2023anomaly}. In contrast, the Secure Water Treatment (SWaT) dataset \cite{mathur2016swat}, which includes both physical and network data, proves ideal for our experiments due to its comprehensive coverage.

We utilize the SWaT dateset, which is collected at the Singapore University of Technology from a water treatment system. The SWaT dataset includes uninterrupted 11-day operational period data, during which the system runs from an empty state to a fully operational state. For the initial seven days, the system operates under normal conditions without any attacks or malfunctions. In the remaining four days, while data collection continues, a range of network and physical attacks are launched on the SWaT system. We are grateful for the open source code provided by S Tuli et al, who propose the TranAD \cite{tuli2022tranad}. This resource is instrumental to adapt the nine distinct baseline models to the SWaT dataset in our experiments. Because our methodology is uniquely capable of detecting multi-domain data simultaneously, we provide data specific to either the physical or network domain in SWaT, as delineated in thier works.

In this dataset, the physical domain includes 51 columns of values, representing the measured values per second of 51 nodes (sensors or actuators). On the other hand, the network domain includes 16 columns of values extracted from the transmitted packets in the ICS network during the measurement period. We first determine that each domain contains 51 feature vectors, representing the features of all nodes in the ICS, which is beneficial for constructing the graph structure. Then, we extract the initial network domain data of each node into three feature values: the number of sent packets, the number of received packets, and the sending overall payload, corresponding to the measured values of each node on the physical domain.

Table 1 summarizes the statistics of the SWaT dataset in both physical and network domains.

\begin{table}[ht]
\setlength{\tabcolsep}{11pt}
\renewcommand\arraystretch{1.8}
\centering
\caption{The statistics of the SWaT dataset in different domains.}
\label{tab_rob}
\begin{tabular}{ccccc}
   \toprule
   \textbf{Domain}  & \textbf{Train} & \textbf{Test} & \textbf{Features}  & \textbf{Anomalies}\\
   \midrule
   Physical  & 21,830 & 34,201 & 51 & 16.61$\%$ \\
   Network & 2,1830 & 34,201 & 3 & 16.61$\%$ \\
   \bottomrule
\end{tabular}
\end{table}

\subsection{Evaluation Method}

The precision, recall, F1-Score and False Positive Rate (FPR) are applied to evaluate the performance of our proposed model. Depending on the optimization direction of the metrics, they are divided into positive (precision, recall, and F1 score) when higher values indicate better performance,  and negative (FPR) when lower values indicate better performance \cite{rainio2024evaluation}. Let TP represent the number of sequences that are correctly predicted as positive, TN denote the number of sequences that are correctly classified as negative, FN denote the number of traces that are positive but are incorrectly predicted as negative, and FP indicate the number of traces that are negative but are predicted as positive. The calculation methods for these metrics are shown in Eq. (16)-(19). To detect anomalies, we use the maximum anomaly score on the validation dataset to set a threshold. During testing, any time step where the anomaly score exceeds the threshold will be considered abnormal.

\begin{equation}
 Precision = {{TP} \over {TP + FP}}
\end{equation}

\begin{equation}
 Recall = {{TP} \over {TP + FN}}
\end{equation}

\begin{equation}
 F1-Score = {{2*Precision*Recall} \over {Precision + Recall}} 
\end{equation}

\begin{equation}
 FPR = {{FP} \over {FP + TN}}
\end{equation}

\subsection{Results and Discussion}\label{subsec:results_discussion}

Our model outperforms when evaluated against nine other baseline methods, as well as in Ablation Experiments. Through evaluation with the baselines, our model not only achieves the best precision, FPR and F1-score, demonstrating its outstanding ability to accurately identify anomalies but also an excellent balance of precision and recall. These excellent performances are crucial for the practical application of our model in ICSs. Additionally, the ablation experiments show that both the attention mechanism and the multi-gradient descent optimization algorithm are indispensable components of our model. These findings suggest that the attention mechanism and the multi-gradient descent optimization algorithm all contribute to model performance, and our model is accurate and reliable for real-world ICS anomaly detection. In this section, we provide detailed descriptions of the results and in-depth discussions.%main results

\begin{table*}[ht]
\setlength{\tabcolsep}{38pt}
\renewcommand\arraystretch{1.8}
\centering
\caption{Anomaly detection performance comparison in terms of FPR($\%$), percision($\%$), recall($\%$), and F1-score($\%$) of our model with baseline methods.}
\label{tab_compar}
\begin{tabular}{ccccc}
    \toprule  
    \textbf{Method}%第一道横线
	% \multirow{2}{*}{\textbf{Method}}&                      %合并两行居中显示
	% \multicolumn{3}{c}{\textbf{SWaT}}\cr 
    % \cmidrule{2-4}                %\cmidrule分隔线，可以指定在列
	&\textbf{FPR} & \textbf{Precision} & \textbf{Recall} & \textbf{F1}\\
	\midrule                                      %第二道横线 
	%Anomal-E &0 &0 &0\\
	DTAAD &13.33 &59.88 &99.99 &74.90\\
	GDN &10.70 &64.91 &99.45 &78.55\\
	LSTM-AD &13.33 &59.88 &99.99 &74.90\\
    MAD-GAN &13.57 &59.45 &99.99 &74.57\\
    MSCRED &13.33 &59.89 &99.99 &74.91\\
    MTAD-GAT &13.39 &59.78 &99.99 &74.83\\
    OmniAnomaly &13.36 &59.83 &99.99 &74.87\\
    TranAD &13.35 &59.85 &99.99 &74.88\\
    USAD &13.26 &60.02 &99.99 &75.01\\
    \textbf{Our Model} &\textbf{3.07} &\textbf{84.65} &\textbf{85.12} &\textbf{84.88}\\
	\bottomrule                                   
\end{tabular}%
\end{table*}

\subsubsection{Comparison with Baselines}

To demonstrate the overall performance of our model, we compare it with nine methods for the detection of multivariate time series anomalies in ICSs. Table \ref{tab_compar} provides the FPR, precision, recall, and F1 scores for our model and baseline models for the SWaT dataset. As shown in Table \ref{tab_compar}, our model shows excellent generalization capability and achieves the best FPR and F1 scores consistently on the SWaT dataset. This enhancement is due to our model's cross-domain representation learning mechanism, which integrates global and local information of nodes in multiple domains and enables the model to discover more potential nonlinear relationships. The shared embeddings take into account both single-domain and cross-domain semantics and provides broader node and correlation information for further single-domain representation learning.

\textbf{Our models’ performance excels several state-of-the-art deep learning-based single-dimensional anomaly detection models (e.g., TranAD, GDN, LSTM-AD an,d so on), with an F1 score gain of no less than 6.33$\%$.} Our model further introduces network and other domain information into the time series information of general physical sensors for learning and training. Through cross-domain representation learning, multi-domain information is first fused for learning, and cross-domain embedding is split into physical and other domains for specific learning, ultimately obtaining node embeddings of ICSs on multiple domains. Therefore, compared to the single-domain detection model, our model can obtain the best performance.

\textbf{Our model surpasses the other multidimensional anomaly detector (e.g., MTAD-GAT) evaluated in this paper in terms of all evaluation metrics.} MTAD-GAT considers each univariate time-series as an individual feature and tries to model the correlations between different features explicitly, while the temporal dependencies within each time series are modeled at the same time. MTAD-GAT leverages two parallel graph attention layers to learn the relationships between different time-series and timestamps dynamically, while our model considers more generalized multidimensional information, where data from different domains is measured and represented independently. In addition, our model adopts an aggregation-splitting training framework, which helps to mine more cross-domain information and assist in feature learning on various single domains. Profiting from the multi-graph representation method, our model can comprehensively and efficiently learn the information of nodes on multiple dimensions simultaneously during the cross-domain representation learning stage and explore the potential correlations between different domains of nodes, providing more prior knowledge for specific domain learning and more selectivity for anomaly detection. Therefore, our model achieves the best evaluation performance compared to other multidimensional anomaly detection models.
 
\textbf{Compared with all other evaluated anomaly detection models, our model exhibits the best performance, with the highest F1 score and precision and the smallest difference between precision and recall, indicating that our model has the best performance balancing ability.} Although some methods (e.g., TranAD, MSCRED, and DTAAD) can obtain higher scores than the proposed model on recall metrics, they are worse on other metrics. This is due to the fact that most of these models only use sensor values from 51 nodes as training data and implement deep learning-based anomaly detection, thus making it difficult to mitigate the sparsity problem of ICS sensor data. Although some methods have relatively high precision, the recall is generally low and do not achieve relatively balanced detection results. This suggests that the relevant information is not fully utilized and is subject to some problems such as false alarm, inapplicability to minority issues, and insensitivity to changes in practical situations\cite{chicco2020advantages}.

\begin{figure}[!t]
    \centering
    \includegraphics[width=3.5in]{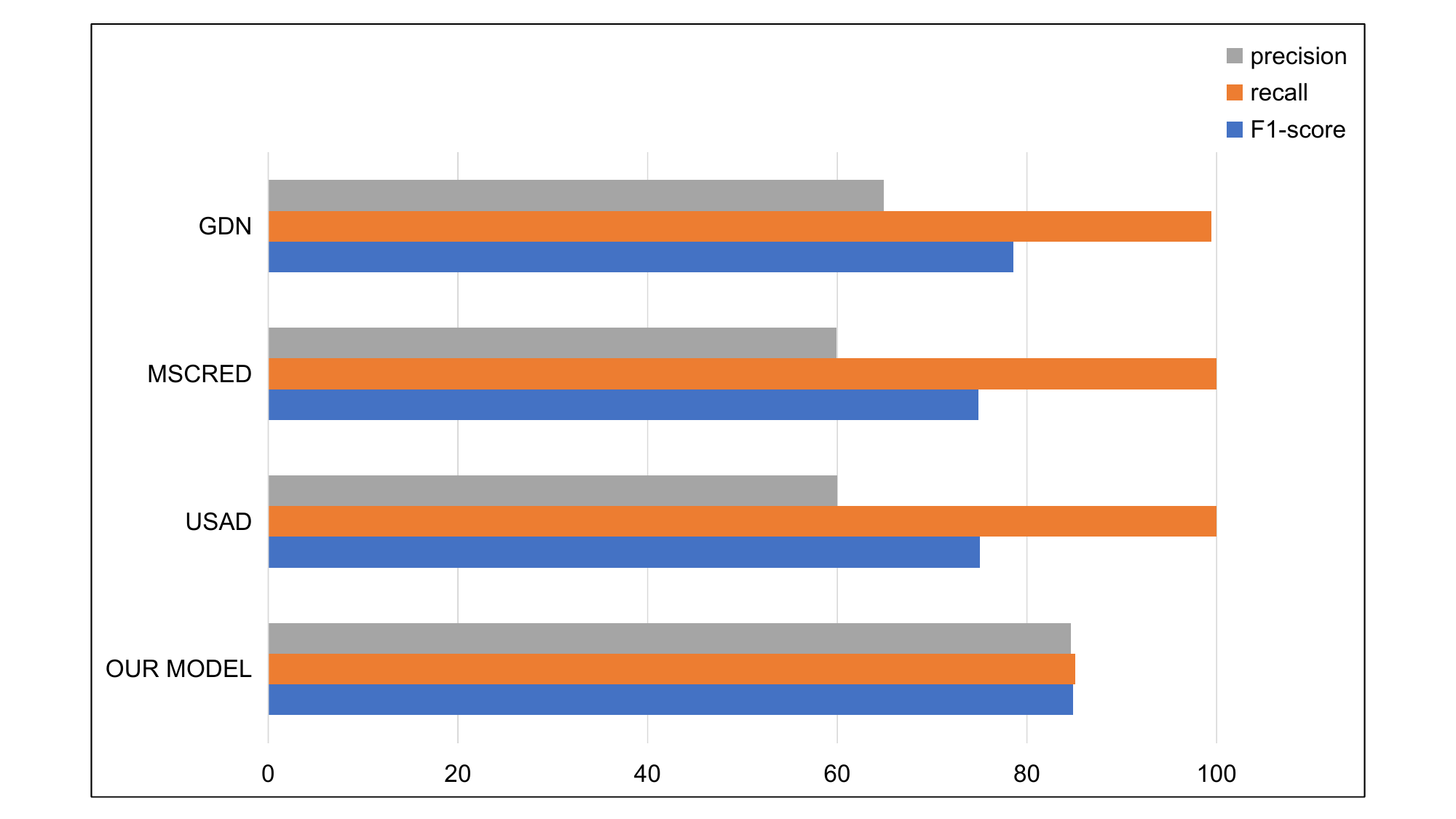}
    \caption{Comparison of performance balance among our model, GDN, USAD, and MSCRED.}
    \label{fig:fig_4}
\end{figure}

Although our model does not achieve the highest precision or recall, its evaluation results are more balanced, with both metrics tending towards higher values. Figure \ref{fig:fig_4} shows the comparison between the three models that achieved the best values in all indicators except for our model in the experiment and our model, making it easy to visually observe the difference in size of the indicators. There are significant differences between the precision and recall performance of the GDN, USAD, and MSCRED models, while our model shows an average level of excellence in various indicators and reaches the best level of surpassing other baselines on the F1 Score.

To comprehensively and reliably demonstrate the advantages of our model in practical applications, anomaly detection performance is evaluated against baseline models using the FPR. This metric not only illustrates the model's efficiency in correctly identifying true anomalies but also highlights its superior ability to minimize false alarms, providing a robust indication of its practical utility in real-world scenarios. As shown in Table \ref{tab_compar}, our model achieves an excellent result of 3.07$\%$ on the metric FPR, which is the lowest compared with the baselines, indicating superior performance in reducing false positives. The results show that our model is more advantageous in scenarios where the cost of false positives is serious, and lower false positives are beneficial to maintain the normal operation of ICS and ensure the reliability of anomaly detection results \cite{mijalkovic2022reducing}.

\subsubsection{Ablation Experiments}
In this section, we evaluate the effectiveness of some methods (e.g., the attention mechanism, etc.) in our model by comparing F1 scores, precision, and recall with ablation experiments.

\textbf{Effectiveness of Graph Attention.}
We examine the influence of the graph attention mechanism in our model by disabling it and instead aggregating using equal weights assigned to all neighbors. From Table \ref{tab_4}, we can find that the removal of the graph attention mechanism causes a 6.49$\%$ decline in the average F1 score. Since nodes in ICSs have very different behaviors, treating all neighbors equally makes noise and misleads our model. This verifies that employing the attention mechanism in our model can better learn complex graph structure information, enable the model to better mine node features based on correlation, and thus perform more accurate anomaly detection.

\begin{table}[ht]
\setlength{\tabcolsep}{17pt}
\renewcommand\arraystretch{1.8}
\centering
\caption{Anomaly detection accuracy in terms of percision ($\%$), recall ($\%$), and F1-score ($\%$) of our model and its variants.}
\label{tab_4}
\begin{tabular}{cccc}
    \toprule     %第一道横线
    \textbf{Method}%第一道横线
	% \multirow{2}{*}{\textbf{Method}}&                      %合并两行居中显示
	% \multicolumn{3}{c}{\textbf{SWaT}}\cr 
 %    \cmidrule{2-4}                %\cmidrule分隔线，可以指定在列
	& \textbf{Precision} & \textbf{Recall} & \textbf{F1}\\
	\midrule                                      %第二道横线 
	%Anomal-E &0 &0 &0\\
    \textbf{Our Model} &\textbf{84.65} &\textbf{85.12} &\textbf{84.88}\\
    \quad-Attention &99.25 &64.78 &78.39\\
	\bottomrule                                   %第三道横线
\end{tabular}%
\end{table}

Vanilla GCNs utilize a fixed convolutional kernel to aggregate information from neighboring nodes, attributing equal influence to all \cite{kipf2016semi}. However, in multi-dimensional data contexts, the significance of different neighboring nodes can vary substantially, which Vanilla GCNs overlooks. Moreover, learning node features in high-dimensional spaces presents substantial complexity, with abundant potential node association information yet to be explored. Integrating an attention mechanism into the vanilla GCNs framework allows for the dynamic weighting of neighbor nodes based on their importance, enabling the model to flexibly aggregate information and capture more intricate inter-node relationships \cite{gupta2023knowledge}. Without the attention mechanism, the effectiveness of the model in information aggregation and feature selection diminishes, leading to an inability to fully capture all pertinent positive samples. Consequently, the model becomes more conservative, resulting in an increase in precision but a notable decrease in recall.

\textbf{Effectiveness of the Multi-Gradient Descent Optimizer.}
To verify the effectiveness of employing a multi-gradient descent optimizer in our model, we replace the multi-gradient descent optimization algorithm with a weighted sum method that statically calculates all task losses and sets the static ratio of loss for two domain-specific learning tasks: physical weight 0.5 vs. network weight 0.5, physical weight 0.25 vs. network weight 0.75, and physical weight 0.75 vs. network weight 0.25. We record the evaluation results (F1-score, precision, and recall) of the above three cases and the original model, whose results are shown in Table \ref{tab_5}.

\begin{table}[ht]
\setlength{\tabcolsep}{17pt}
\renewcommand\arraystretch{1.8}
\centering
\caption{Performance comparison of our model with models employing the static calculation loss sum method instead of the multi-gradient descent optimizer method.}
\label{tab_5}
\begin{tabular}{cccc}
    \toprule            
    \textbf{Method}%第一道横线
	% \multirow{2}{*}{\textbf{Method}}&                      %合并两行居中显示
	% \multicolumn{3}{c}{\textbf{SWaT}}\cr 
 %    \cmidrule{2-4}                %\cmidrule分隔线，可以指定在列
	& \textbf{Precision} & \textbf{Recall} & \textbf{F1}\\
	\midrule                                      %第二道横线 
	%Anomal-E &0 &0 &0\\
	0.5 : 0.5 &83.15 &82.34 &83.87\\
	  0.25 : 0.75 &49.30 &84.46 &62.28\\
    0.75 : 0.25 &80.99 &83.39 &82.79\\
    \textbf{Our Model} &\textbf{84.65} &\textbf{85.12} &\textbf{84.88}\\
	\bottomrule                                   %第三道横线
\end{tabular}%
\end{table}

The evaluation result shows that setting the loss ratio of the physical domain and the network domain to 0.5 vs. 0.5 can achieve better results than other static weighting methods, but our current model using the multi-gradient descent optimizer is still better than at least 1.01$\%$ in the F1 score. Therefore, the introduction of a multi-gradient descent optimization algorithm is beneficial for the model to better dynamically adjust the weights of multi-task learning, providing optimization guarantees for accelerating model convergence and improving model performance. Furthermore, we find that static loss calculation methods are difficult to explain and obtain the optimal weight allocation parameters, making it difficult to effectively and reasonably solve competitive multi-objective optimization problems.

\iffalse
\begin{figure}
    \centering
    \begin{minipage}[t]{0.5\linewidth}
        \centering
        \includegraphics[width=\textwidth]{7-1.jpg}
        \centerline{(a) Recall  }
    \end{minipage}%
    \begin{minipage}[t]{0.5\linewidth}
        \centering
        \includegraphics[width=\textwidth]{7-2.jpg}
        \centerline{(b) Precision }
    \end{minipage}

    \begin{minipage}[t]{0.5\linewidth}
        \centering
        \includegraphics[width=\textwidth]{7-3.jpg}
        \centerline{(c) F1 score    }
    \end{minipage}%
    \begin{minipage}[t]{0.5\linewidth}
        \centering
        \includegraphics[width=\textwidth]{7-4.jpg}
        \centerline{(d) ROC }
    \end{minipage}
    
    \caption{Comparison with different number of behavior units.}
    domal{f:comparison-number}
\end{figure}

\begin{figure}
    \begin{minipage}[t]{0.5\linewidth}
        \centering
        \includegraphics[width=\textwidth]{8-1.jpg}
        \centerline{(a) Recall }
    \end{minipage}%
    \begin{minipage}[t]{0.5\linewidth}
        \centering
        \includegraphics[width=\textwidth]{8-2.jpg}
        \centerline{(b) Precision }
    \end{minipage}

    \begin{minipage}[t]{0.5\linewidth}
        \centering
        \includegraphics[width=\textwidth]{8-3.jpg}
        \centerline{(c) F1 score }
    \end{minipage}%
    \begin{minipage}[t]{0.5\linewidth}
        \centering
        \includegraphics[width=\textwidth]{8-4.jpg}
        \centerline{(d) ROC}
    \end{minipage}
    \caption{ Comparison with different lengths of sequences. }
    \label{f:comparison-length}
\end{figure}

\fi

\subsubsection{Sensitivity Analysis}

We evaluate our model's sensitivity to different learning rate settings and combinations across multiple tasks. The learning rates for the three different tasks are set, and we observe the model's performance in anomaly detection. LR1 and LR2 represent the learning rates for the physical domain learning and network domain learning tasks, respectively, while LR3 represents the learning rate for the cross-domain learning task. This sensitivity is evaluated by comparing the Precision, Recall, and F1-score metrics.

\begin{table}[h!]
\setlength{\tabcolsep}{9.2pt}
\renewcommand\arraystretch{1.8}
\centering
\caption{Performance comparison of different learning rate settings (All values in $\%$).}%单位都是百分号
\begin{tabular}{cccccc}
\toprule
\multicolumn{3}{c}{\textbf{Learning Rates}} & \multirow{2}{*}{\textbf{Precision}} & \multirow{2}{*}{\textbf{Recall}} & \multirow{2}{*}{\textbf{F1-Score}} \\
\cmidrule(lr){1-3} 
\textbf{LR1} & \textbf{LR2} & \textbf{LR3} &  &  &  \\
\midrule
0.001 & 0.001 & 0.001 & 99.99 & 64.24 & 78.24 \\
0.001 & 0.001 & 0.1 & 99.99 & 64.24 & 78.24 \\
0.001 & 0.1 & 0.1 & 85.01 & 84.72 & 84.86 \\
0.1 & 0.1 & 0.1 & 82.25 & 83.17 & 83.44 \\
0.1 & 0.1 & 0.001 & 81.07 & 83.65 & 82.75 \\
\textbf{0.1} & \textbf{0.001} & \textbf{0.001} & \textbf{84.65} & \textbf{85.12} & \textbf{84.88} \\
\bottomrule
\end{tabular}
\end{table}

As shown in Figure V, the results indicate that the model's performance is sensitive to the learning rate settings for different tasks. Specifically, setting LR1, LR2, and LR3 to identical values hinders the model from achieving optimal anomaly detection performance. This is due to the varying complexity and convergence rates of each task during the learning process. When the other two learning rates are held constant, increasing the value of LR1 affects the F1-score by up to 6.64$\%$, indicating a high sensitivity to LR1. This sensitivity can be attributed to the high feature complexity associated with the physical domain learning task that LR1 regulates \cite{crawshaw2020multi}. The optimal combination, LR1 = 0.1, LR2 = 0.001, and LR3 = 0.001, achieves a balance between Precision and Recall, enhancing overall performance. Consequently, this configuration was selected as the final setting for our model.

\section{Conclusion}\label{s:conclusion}
This paper presents an anomaly detection approach based on cross-domain representation learning, which combines ICS data on multiple domains for cross-domain learning and anomaly detection. We propose a multi-graph representation method to uniformly represent ICS multi-domain data on just one graph structure and then design an attention-based cross-domain graph convolutional network for learning embedding. We evaluate our model on a large-scale real-world dataset, and experimental results show that our model outperforms baselines. In addition, our model can better balance the relationship between reducing false positives and improving anomaly detection precision, providing a more practical and ideal ICS anomaly detection model. Future works may come from two aspects. First, more practical ICS anomaly detection datasets with aligned multi-domain data will be explored, and our model will be developed to adapt to a wider range of application scenarios. Secondly, the advantages and costs of cross-domain learning in ICS anomaly detection will be further explored.

\section*{Acknowledgments}

This work was supported by the National Natural Science Foundation of China under Grants No. 62302122 and No. 62172123, the Natural Science Foundation of Heilongjiang Province of China under Grants No. LH2023F017, and CCF-Huawei Populus Grove Fund under Grants No. CCF-HuaweiSY202411.

\vspace{11pt}

\begin{IEEEbiography}[{\includegraphics[width=1in,height=1.25in,clip,keepaspectratio]{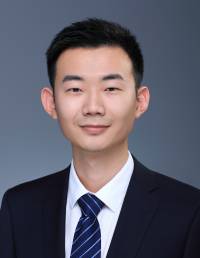}}]{Dongyang Zhan}
is an associate professor in School of Cyberspace Science at Harbin Institute of Technology. He received the B.S. degree in Computer Science from Harbin Institute of Technology from 2010 to 2014. From 2015 to 2019, he has been working as a Ph.D. candidate in School of Computer Science and Technology at HIT. His research interests include cloud computing and security.
\end{IEEEbiography}

\vspace{11pt}

\begin{IEEEbiography}[{\includegraphics[width=1in,height=1.25in,clip,keepaspectratio]{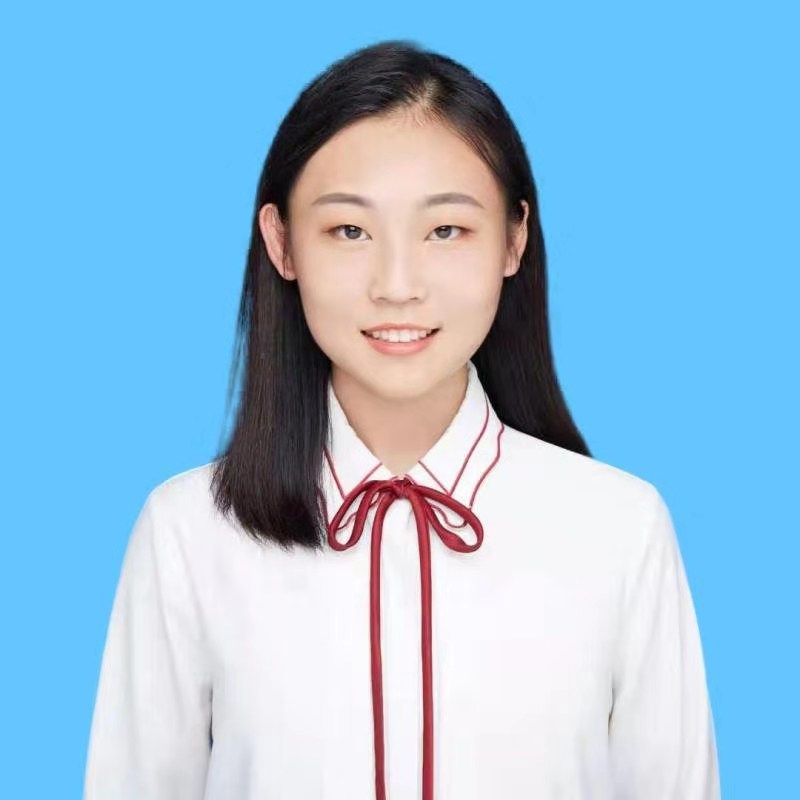}}]{Wenqi Zhang}
is a master's student from the Harbin Institute of Technology, China. Her research focuses on IoT security.
\end{IEEEbiography}

\vspace{11pt}

\begin{IEEEbiography}[{\includegraphics[width=1in,height=1.25in,clip,keepaspectratio]{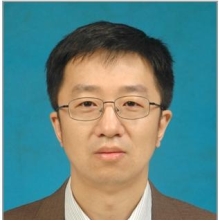}}]{Xiangzhan Yu}
is a professor in School of Cyberspace Science at Harbin Institute of Technology. His main research fields include: network and information security, security of internet of things and privacy protection. He has published one academic book and more than 50 papers on international journals and conferences.
\end{IEEEbiography}

\vspace{11pt}

\begin{IEEEbiography}[{\includegraphics[width=1in,height=1.25in,clip,keepaspectratio]{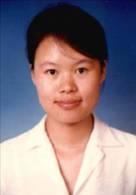}}]{Hongli Zhang}
received her BS degree in Computer Science from Sichuan University, Chengdu, China in 1994, and her Ph.D. degree in Computer Science from Harbin Institute of Technology (HIT), Harbin, China in 1999. She is currently a Professor in School of Cyberspace Science in HIT. Her research interests include network and information security, network measurement and modeling, and parallel processing.
\end{IEEEbiography}

\vspace{11pt}

\begin{IEEEbiography}[{\includegraphics[width=1in,height=1.25in,clip,keepaspectratio]{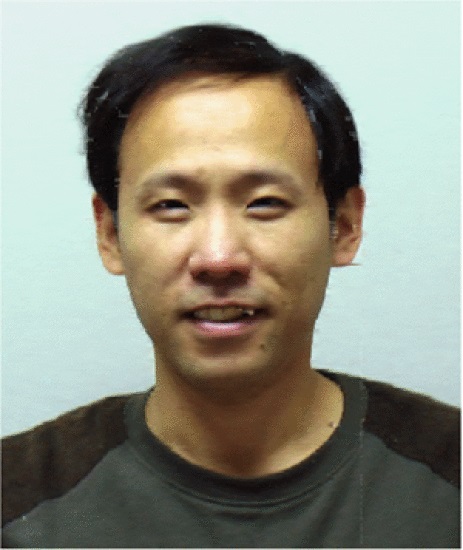}}]{Lin Ye}
received the Ph.D. degree at Harbin Institute of Technology in 2011. From January 2016 to January 2017, he was a visiting scholar in the Department of Computer and Information Sciences, Temple University, USA. His current research interests include network security, peer-to-peer network, network measurement and cloud computing.
\end{IEEEbiography}

\begin{IEEEbiography}[{\includegraphics[width=1in,height=1.25in,clip,keepaspectratio]{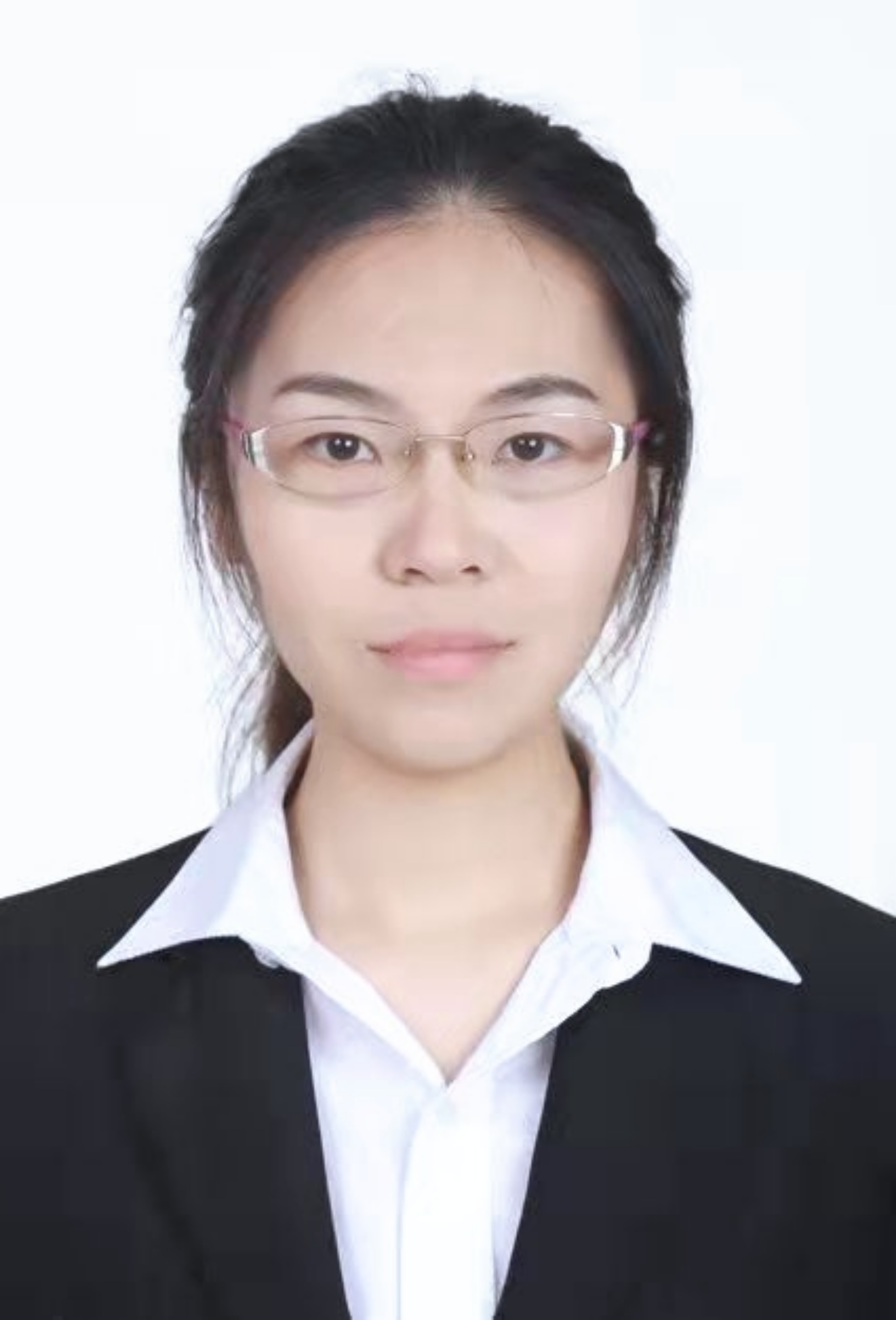}}]{Zheng He}
is an engineer in Heilongjiang Meteorological Bureau. She received her bachelor's and Master’s degrees in Meteorology Science in Nanjing University of Information Science and Technology from 2011 to 2018. From 2018, she has been working in Weather Modification Office of Heilongjiang Province. Her research interests include climate change, weather modification and machine learning.
\end{IEEEbiography}

\vfill

\end{document}